\documentclass[12pt]{iopart}
\usepackage{color}
\usepackage{graphicx}
\usepackage{amsmath, amsthm, amssymb}
\usepackage{latexsym,amsfonts,color,amsthm}
\usepackage[normalem]{ulem}
\usepackage{citesort}
\usepackage{tabularx}
\usepackage{mathtools}
\usepackage{enumerate}
\usepackage[square, numbers, comma, sort&compress]{natbib}

\def\beq{\begin{equation}}
\def\eeq{\end{equation}}

\begin{document}

\title{Clustering in  two models of interacting motors} 
\author{Jim Chacko$^{1~*}$, Sudipto Muhuri$^{2~\dag}$ and  Goutam Tripathy$^{3,4~\#}$} 

\address{$^1$ Christian College, Angadical South PO, Chengannur,  Kerala 689122, India.}
\address{$^2$ Department of Physics, Savitribai Phule University of Pune, Pune Maharashtra 411 007, India.}
\address{$^3$ Institute Of Physics, Sachivalaya Marg, Sainik School PO, Bhubaneswar 751005, India.}
\address{$^4$ Homi Bhaba National Institute, Anushaktigar, Mumbai 400094, India.}
\ead{$^*$~jim.chacko.v@gmail.com, $\dag$~muhuri@physics.unipune.ac.in, $^\#$~goutam@iopb.res.in} 

\date{\today}

\begin{abstract}
We study a two-species bidirectional exclusion process, and a single species variant,  which is
motivated by the motion of organelles and vesicles along
microtubules. Specifically, we are interested in the clustering of the
particles and appearance of a single large cluster as the ratio $Q$ of
the translation to switching rates is varied. We find that, although
for a finite system, there is a clustering phenomenon in which the
probability of finding a single large cluster changes from being
negligible to having finite values, the phenomenon shifts to larger
$Q$ values as the system size is increased. This suggests that the
observed clustering is not a true (non-equilibrium) transition in the
thermodynamic sense but rather a finite-size effect.
\end{abstract}

\maketitle

\section{\label{sec1-Introduction}Introduction}

One-dimensional driven lattice gas models have been the subject of
considerable interest due their remarkably rich stationary and
dynamical behaviour \cite{ref_schutzrev}. Further, the inherent
simplicity of these models are useful for application to diverse
phenomena \cite{ref_traffic, ref_granular} including those in biology
\cite{ref_bio}.

Bidirectional transport of cargo vesicles and organelles along
one-dimensional cytoskeletal filaments has been observed in context of
intracellular transport and studied fairly extensively for the
dynamics of pigment granules in melanophores \cite{ref_borisy},
mitochondria in the axons of neuronal cells \cite{ref_hollenbeck,ref_welte},
endosomes in {\it Dictyostelium} cells \cite{ref_roop} and lipid-droplets
in {\it Drosophila} embryos \cite{ref_lipo,ref_grosslipid}.  Bidirectionality
is achieved by the collective action of oppositely directed motor
proteins (such as kinesin and dynein) which results in the translation
of the cargo along the filament along with the directional switching
of the transported cellular cargo \cite{ref_cell,ref_howard}.

The interplay of the switching dynamics of individual particles and
the collective movement of particles in 1D has been studied
theoretically using a two species lattice gas model in
\cite{ref_madan}. There, it has been shown that if the directional
switching rate ($\alpha$) is much faster than the translation rate
($v$), the steady state density and current profiles of the particles
are homogeneous in the bulk and are well described by a mean field
theory, and the model could be mapped to the exactly solvable
Partially Asymmetric Exclusion Process (PASEP)
\cite{ref_sandow}. Furthermore, it has been argued that the mean field
theory fails away from this fast switching limit, and the steady state
behaviour is characterized as a function of the ratio $Q=v/\alpha$ of
the translation and net switching rates.  In \cite{ref_madan}, it was
observed that in the limit of very slow switching rates (compared to
the translation rates), the system approaches a jammed phase with a
net current that tends to zero as $J\sim 1/Q$. It was further argued
that in many realistic cellular situations involving transport of
cellular cargo, such as transport of lipid-droplets in wild type
  Drosophila embryos \cite{ref_grosslipid}, the switching rates are
much slower than the translation rates so that $Q$ is high. This
would, in turn, imply an enhanced tendency of jamming of these cargo
aggregates and formation of clusters during the transport on the
cellular filament. This naturally invites the question as to how $Q$
affects the cluster size distribution of these cellular cargoes
involved in bidirectional transport. We study this issue within the
framework of a minimal model which is defined on a one-dimensional
lattice and described in Ref.\cite{ref_madan}. However, unlike as in
\cite{ref_madan}, where lattice with open boundary is considered, we
study this model by imposing periodic boundary condition on the
lattice and thereby also fixing the density of vacancies on the
lattice.

We first study the cluster size distribution of particles for the
minimal model using Monte Carlo simulations for a wide range of $Q$.
We look at the cluster size distribution in the steady state and see
the appearance a single large cluster (the aggregate) as the value of
$Q$ is increased value. We study the probability of occurrence of this
large cluster for different system sizes $L$ as well as the density
$\rho$ of particles. Although, the appearance of a large aggregate
seems like a nonequilibrium phase transition, we find that, for a
given $Q$, as the system size is increased the probability of the
large cluster decreases and eventually it disappears.  We further
study another model derived by tweaking the minimal model in which
similar chipping and aggregation mechanisms of particles exist but
with a single species of particles. While it shares qualitative
features of the minimal model, there are quantitative differences.

\section{\label{sec2-2SBEP}Two Species Bidirectional Exclusion Process}

\begin{figure}[h]
\centering
\includegraphics[width=0.5\textwidth]{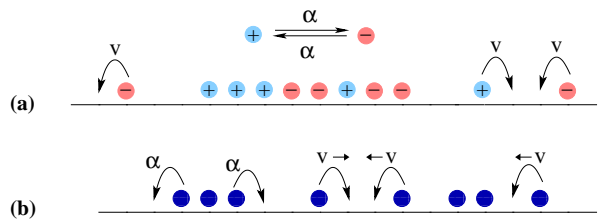}
\begin{center}
\caption{\label{fig3.1} Schematic representation of the dynamical processes for (a)
  the Two Species Bidirectional Exclusion Process (2S-BEP). (b) the Symmetric Exclusion Process with
  Directional Memory (SEP-DM). }
\end{center}
\end{figure}

We mimic the microtubule - molecular motor - vesicle system as a Two
Species Bidirectional Exclusion Process (2S-BEP) in the following
manner.  We represent a single microtubule filament as a
one-dimensional lattice with $L$ sites
(Fig.~\ref{fig3.1}(a)). The
cellular cargo carried by the collective action of motors, is
considered here as a particle; it can either be in the $\oplus$ state
(moving anticlockwise) or in the $\ominus$ state (moving clockwise),
with a fixed rate of switching $\alpha$ between the two states
\cite{ref_note1}.  The switching may be thought of as a result of a
 tug-of-war mechanism by oppositely directed motors or by
external regulation \cite{ref_welte}. The two states of the particles
can, in turn, be regarded as particles of two species.
Each lattice site $i$ is either
vacant or occupied by a particle - either $\oplus$ or $\ominus$. The
$\oplus$ and $\ominus$ translocate to the adjacent vacant site on the
right or left respectively with a rate $v$. In all our simulations, initially,
equal number of particles of both species are taken. The various
dynamical processes are schematically displayed in
Fig.~\ref{fig3.1}(a). The total density of particles (of both types
together) is denoted by $\rho$ and is conserved under the dynamics,
but number/density of each type of particles is not conserved
separately.

We study the 2S-BEP using Monte Carlo simulations. For a given system
size $L$ and overall particle density $\rho$, the initial condition is
a random distribution of $N=L\rho$ particles over $L$ sites. In one
microstep, a site is chosen randomly, and if it is occupied its sign
is changed with rate $\alpha$ and is moved to its appropriate (empty)
neighbouring site with rate $v$. $L$ such microsteps constitute one MC
step. We evolve the system for $T_0$ MC steps to reach steady state
and compute various steady state averages over further $T$ MC
steps. We further, do an ensemble average over $M$ realizations of
initial conditions. Range of values of these quantities in our
simulations are: $T_0 \sim 10^5-10^6$, $T \sim 10^7-10^8$, and $M \sim 10^4-10^5$.

\begin{figure} [htbp]
\begin{center}
\includegraphics[width=0.6\textwidth]{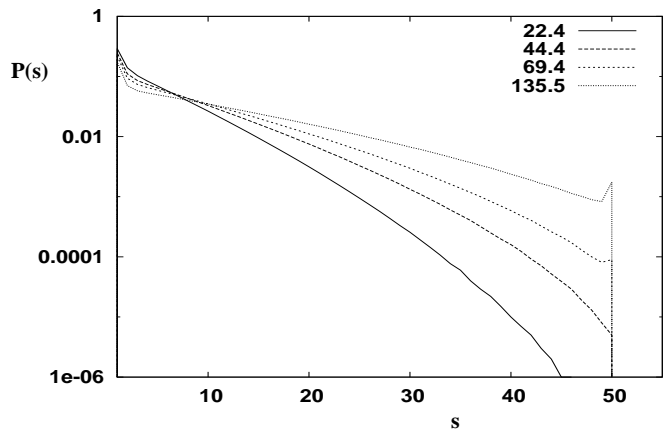}
\end{center}
\caption{\label{fig3.2} Cluster size distribution $P(s)$ for $L=50$ and density $\rho = 0.5$ for different values of $Q$. }
\end{figure}

Particles of each species undergo a totally asymmetric exclusion
dynamics in opposite directions depending on its sign.  A cluster consists of a group of
particles of any sign bounded by two vacant sites and no holes in
between (no distinction is made between the kinds of particles while
determining the size of a cluster).  The smallest blockage is formed if a
$\ldots\oplus\ominus\ldots$ configuration occurs on the lattice - the
only way for it to disappear is by a switching the type of either one
of the particles.  For small values of $Q~O(1)$, both types of particles
move almost freely on the lattice. Although blockages do
form occasionally, the clusters do not grow to large sizes. 

As the value of $Q$ is increased (in our simulations we fix the value
of $v$ and change $\alpha$ to vary $Q$), clusters of different sizes
begin to form. We obtain the probabiity distribution $P(s)$ of
cluster sizes averaged over time $T$ and number of realizations $M$ of
initial conditions. In Fig.~\ref{fig3.2}, we plot the cluster size
distribution $P(s)$ vs $s$ for a number of values of $Q$ for $L=50$
and $\rho=0.5$. We observe that as $Q$ is increased, $P(s)$ broadens
and larger clusters begin to form \cite{ref_note2}. We denote by $P(s_{max})$ the
probability of the largest cluster of particles (by definition the
largest cluster has all the particles, $s_{max}=\rho L$). From
Fig.~\ref{fig3.2}, it appears that beyond a critical value of $Q$,
$P(s)$ becomes nono-monotonic and a local peak appears in the $P(s)$
plot at the largest cluster size $s_{max}$.  This raises the question
whether the appearence of the largest cluster is a (nonequilibrium)
transition.  Below we present evidence that this is not a transition
in the usual thermodynamic sense.

\begin{figure} [htbp]
\begin{center}
\includegraphics[width=0.6\textwidth]{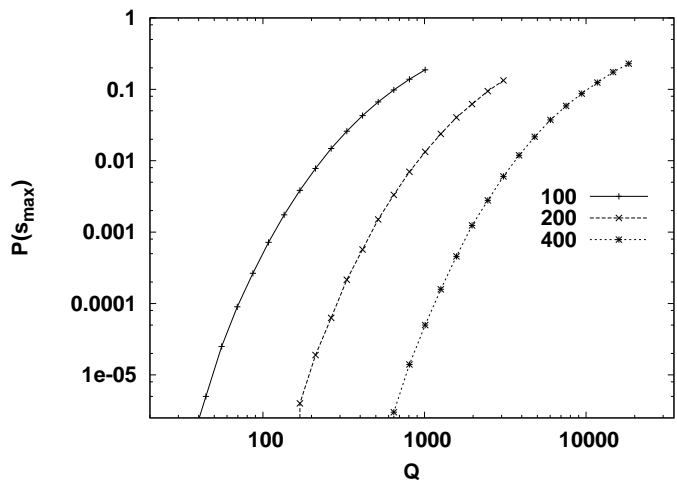}
\caption{\label{fig3.3} $P(s_{max})$ vs $Q$ for $L = 100, 200$ and $400$ and $\rho = 0.5$ for the 2S-BEP.}
\end{center}
\end{figure}

In Fig.~\ref{fig3.3}, we plot variation of $P(s_{max})$ as a function of
$Q$ for a given density $\rho = 0.5$ and for three system sizes
$L=100,200,400$.  Although, for each system size it appears that the
single large cluster occurs for $Q$ beyond a critical value $Q_c$,
i.e., $P(s_{max})=0$ for $Q<Q_c$. However, upon longer steady state
averaging the value of $Q_c$ appears to be lowered.  More importantly,
as the system size is increased, it is seen that, for a given $Q$, the
value of $P(s_{max})$ decreases.  I.e., in the thermodynamic limit
$L\rightarrow\infty$, $P(s_{max})\rightarrow 0$.  Thus, the phenomenon
of formation of the single aggregate as a function of $Q$ is not a
true transition in the thermodynamic sense.  

\begin{figure} [!htbp]
\begin{center}
\includegraphics[width=0.6\textwidth]{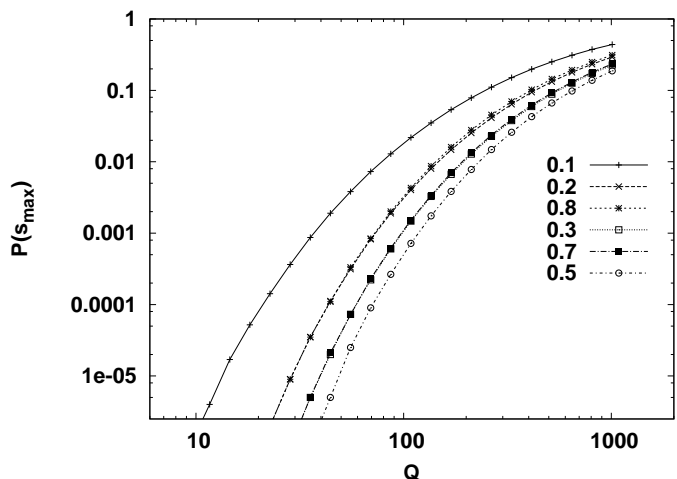}
\vspace{-0.5cm}
\caption{\label{fig3.4} $P(s_{max})$ vs $Q$ for $L = 100$ and total density of particles
$\rho = 0.1, 0.2, 0.3, 0.5, 0.7$ and $0.8$ for the 2S-BEP. The 
$\rho \leftrightarrow  1-\rho$ symmetry is to be noted.
}
\end{center}
\end{figure}

We also compute the variation of $P(s_{max})$ as a function of $Q$ for a
given system size $L=100$ and for a number of densities $\rho$ and the
data is shown in Fig.~\ref{fig3.4}. We notice a $\rho\leftrightarrow
(1-\rho) $ symmetry of the plots. This is due to the fact that $P(s_{max})$ is the same
for the cluster size distribution of the particles $P(s; \rho)$ and
the vacant sites $P(s_h; \rho)$ since the appearance of the largest
aggregate for both are concomitant.  

\begin{figure} [htbp]
\begin{center}
\includegraphics[width=0.6\textwidth]{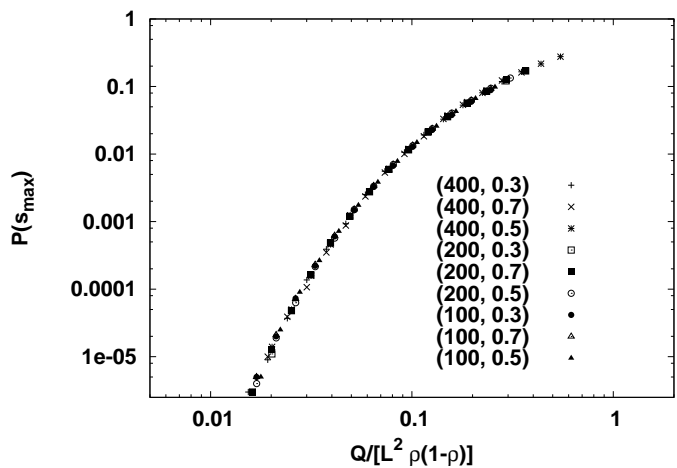}
\caption{\label{fig3.5} Data collapse of $P(s_{max})$ vs $Q/(L^2\rho(1-\rho))$ for the 2S-BEP
for $L = 100, 200$ and $400$ for $\rho = 0.3, 0.5$ and $0.7$.
}
\end{center}
\end{figure}

Although, the clustering seen for large Q is not a true critical transition, nevertheless, we notice an  interesting 
scaling in the nature of  $P(s_{max})$ vs $Q$ plots for
different $L$ and $\rho$ values as follows,
\begin{equation}
P(s_{max}) = F[\frac{Q}{L^\mu\rho(1-\rho)}]; ~~~\mu=2.
\label{eq:psmax}
\end{equation}
The data collapse via Eq.~(\ref{eq:psmax}) of the data shown in Figs.~\ref{fig3.3},~\ref{fig3.4}
is shown in Fig.~\ref{fig3.5}.

\begin{figure} [!htbp]
\begin{center}
\includegraphics[width=0.5\textwidth]{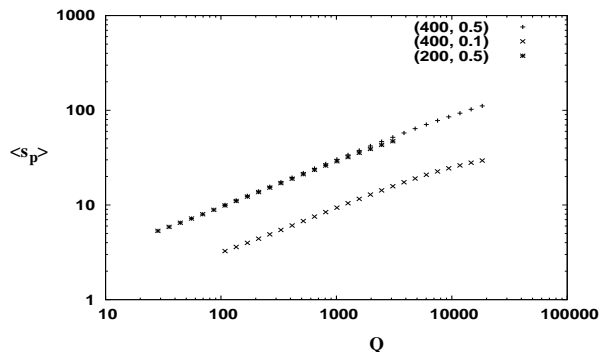}
\caption{\label{fig3.6} The mean cluster size of particles $\langle s_p\rangle$ vs $Q$ for 
  $(L,\rho) = (400, 0.5), (400, 0.3)$ and $(200, 0.5)$ for the 2S-BEP.
  $\langle s_p\rangle \sim \sqrt{Q}$ for the 2S-BEP.
}
\end{center}
\end{figure}

We also computed the mean cluster size for the particles and find that
$\langle s_p\rangle \sim \sqrt{Q}$ for smaller values of $Q<<L^2$
(Fig.\ref{fig3.6}). The mean cluster size for the vacant sites
$\langle s_h\rangle \sim \sqrt{Q}$ as well. As $Q$ increases, the mean
particle and vacancy cluster size increase and saturate at $L\rho$ and
$L(1-\rho)$ respectively, as $Q\rightarrow\infty$, corresponding to
the single largest cluster.  These observations can be combined in to
a single scaling function for the mean cluster size as
\begin{equation}
\langle s_p \rangle = f(\rho) ~Q^\phi ~Y\left(\frac{Q}{g(\rho)L^{1/\phi}}\right); ~~~~~~~Y(u\rightarrow 0)=1,
~~~~Y(u\rightarrow\infty)=u^{-\phi}.
\label{eq:meancs}
\end{equation}
To ensure that the asymptotic $\langle s_p\rangle =L\rho$, we must have $f(\rho)[g(\rho)]^\phi=\rho$.
The data of Fig.~\ref{fig3.6} is consistent with the value of the scaling exponent $\phi=1/2$.
We further conjecture that the two exponents $\mu$ and $\phi$ are indeed related as 
\begin{equation}
\phi=\frac{1}{\mu}.
\label{eq:muphi}
\end{equation}

\section{\label{sec3-SEPDM}Symmetric Exclusion Process with Directional Memory}

\indent The 2S-BEP discussed in the previous section is a two-species model with
particle moves specified by two rates, $v$ and $\alpha$. It displays a
clustering of particles when the ratio $Q = v/\alpha$ is increased.
$P(s_{max})$, the probability to find the largest cluster, has a
$\rho \longleftrightarrow (1-\rho)$ symmetry as well as a system size
dependence. The clusters in the two-species model  grow or fragment
when a $+ (-)$ species switches to a $- (+)$ and hops to the left (right)
until it attaches to the right (left) of another cluster (or itself, in 
the case when only one large cluster exists). Retaining this feature, i.e.,  
the directionality of motion of the particles that chips off from a cluster,
we  introduce a single species model, namely, the Symmetric Exclusion
Process with a Directional Memory (SEP-DM).

\begin{figure} [htbp]
\begin{center}
\includegraphics[width=0.6\textwidth]{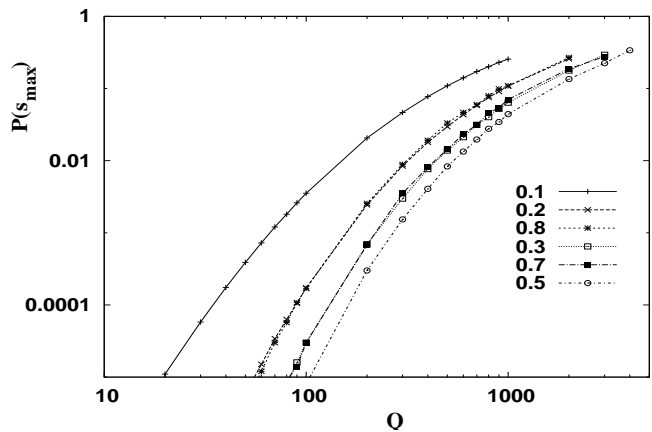}
\caption{\label{fig3.7} $P(s_{max})$ vs $Q$ for SEP-DM for a system of size $L=100$ and 
different density of particles $\rho = 0.1, 0.2, 0.3, 0.5, 0.7$ and $0.8$.
Note the $\rho \leftrightarrow  1-\rho$ symmetry.
}
\end{center}
\end{figure}
The model is defined on a 1D lattice of $L$ sites with periodic
bondary conditions (Fig.~\ref{fig3.1}b).  The dynamics of the
particles are as follows. A particle chips off from the left or right
end of a cluster with a rate $\alpha$ and keeps moving in the same
direction (i.e. the particle breaking off from the left end will keep
moving left/clockwise and one breaking off from the right edge will
keep moving right/anti-clockwise) with a rate $v$ until it attaches
itself to the neighbouring cluster.

In the same spirit as in the case of 2S-BEP, we define $Q = v/\alpha$
to be the ratio of rate of translocation (hopping of single free
particle) to the rate of detachment of a particle from a cluster. We
study the SEP-DM using Monte Carlo simulations. For a given system
size $L$ and overall particle density $\rho$, the initial condition is
a random distribution of $N=L\rho$ particles. We evolve the system for
$T_0$ MC steps to reach steady state and compute various steady state
averages over further $T$ MC steps. We further do an ensemble average
over $M$ realizations of initial conditions. Range of values of these
quantities in our simulations are: $T_0 \sim 10^5-10^6$, $ T \sim
10^7-10^8$ and $M \sim 10^4-10^5$.

\begin{figure} 
\begin{center}
\includegraphics[width=0.47\textwidth]{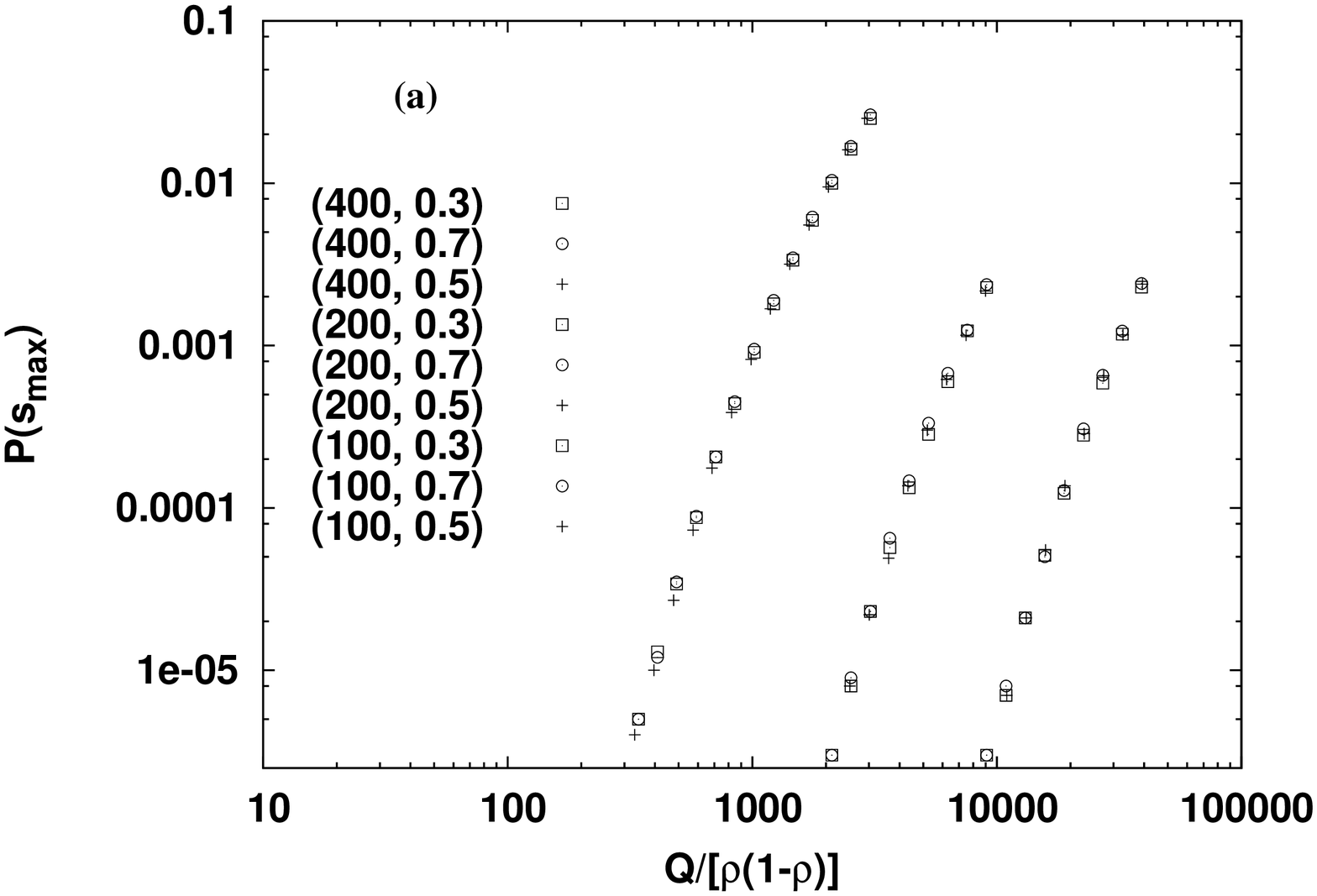}
\includegraphics[width=0.49\textwidth]{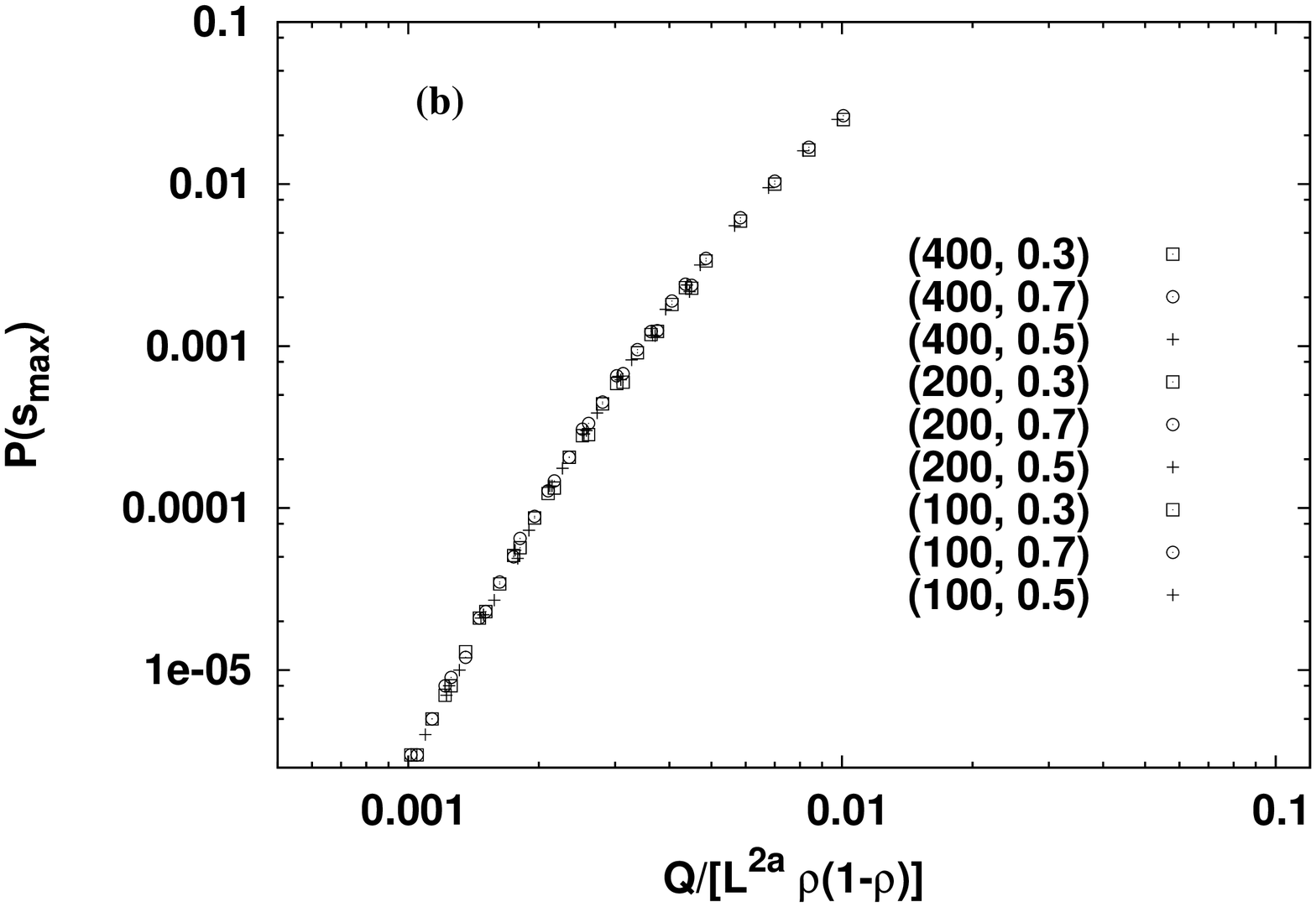}
\caption{\label{fig3.8} {\bf(a)} $P(s_{max})$ vs $(Q/(\rho(1-\rho))$ for the SEP-DM
  for three system sizes: $L=100, 200, 400$ for the densities $\rho =
  0.3, 0.5$ and $0.7$.  {\bf(b)} The
  scaling collapse of $P(s_{max})$ vs $Q$ via Eq.~(\ref{eq:psmax}) but with $\mu=8/3$.}
\end{center}
\end{figure}

Due to the directional memory feature mimicking the movements of two
species of particles in the 2S-BEP, we expect the SEP-DM to have, at
least qualitatively, similar behaviour as the 2S-BEP.  We plot the
probability of the largest cluster $P(s_{max})$ as a function of $Q$
for a number of densities $\rho$ for a system size $L=100$ in
Fig.~\ref{fig3.7}.  The data shows the $\rho\leftrightarrow (1-\rho)$
symmetry also seen in the 2S-BEP model.  In Fig.~\ref{fig3.8}(a), we
plot $P(s_{max})$ vs $\frac{Q}{\rho(1-\rho)}$, taking in to account
this symmetry, for three different system sizes $L$. It is seen from
this plot that as $L$ increases $P(s_{max})$ decreases for the same
value of $Q$. This implies that the appearance of the single cluster
is not a true transition in the case of the SEP-DM as well, as may be
expected.

It turns out that a scaling collapse of the $P(s_{max})$ vs $Q$ data
for different $L$ and $\rho$, values is possible with the scaling form
given in Eq.~(\ref{eq:psmax}) but with $\mu=8/3$. This is shown in
Fig.~\ref{fig3.8}(b). 

We also computed the mean cluster size of particles for SEP-DM as a function of $Q$ and find that
the data is consistent with$\langle s_p\rangle \sim Q^{3/8}$ for smaller values of $Q$
(Fig.\ref{fig3.6}). As $Q$ increases, the mean
cluster size increases and saturate at $L\rho$, as $Q\rightarrow\infty$, corresponding to
the single largest cluster.  Thus, the $L$ and $\rho$ dependence of the mean cluster size
can be put in the scaling form of Eq.~(\ref{eq:meancs}) with $\phi=3/8$. It is interesting to 
note that the relationship Eq.(\ref{eq:muphi}) between the two exponents $\mu$ and $\phi$ holds 
in this case as well.

\begin{figure} [!htbp]
\begin{center}
\includegraphics[width=0.5\textwidth]{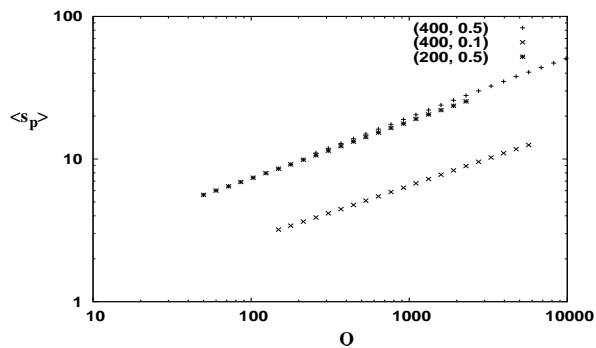}
\caption{\label{fig3.9} The mean cluster size of particles $\langle s_p\rangle$ vs $Q$ for 
  $(L,\rho) = (400, 0.5), (400, 0.1)$ and $(200, 0.5)$ for the SEP-DM.
  The slope of the plots for smaller $Q$'s is consistent with a value of $3/8$.
}
\end{center}
\end{figure}

\section{\label{sec6-Conc}Conclusion}

In this paper, we have studied a two-species bidirectional exclusion
process which was motivated by the motion of organelles and vesicles
along microtubules. Specifically, we were interested in the clustering
of the particles and appearance of a single large cluster as the ratio
$Q$ of the translation to switching rates is varied. We found that
although for a finite system, there appears to be a clustering
phenomena in which the probability of finding the single large cluster
changes from being negligible to having appreciable finite values, the
phenomena shifts to larger $Q$ values as the system size was
increased. This suggests that the observed transition is not a true
transition in the thermodynamic sense but rather a finite-size effect.

We further studied a single species version of the above model in
which the directional memory of the detached particle is
incorporated. This model also shows qualitatively similar clustering
phenomena which vanishes in the thermodynamic limit. 

In both models, the probability of occurence of the largest cluster
and mean cluster size as functions of $Q$ shows nontrivial dependence
on system size $L$ and overall density $\rho$ which can be
characterized by scaling functions with corresponding
exponents. Analytical understanding of these scaling forms and the
values of the scaling exponents remain open for further
investigations.

We would like to comment on a few other one dimensional (1D) models
where such apparent transitions as well as finite-size effects have
been reported. Consider the symmetric conserved mass aggregation model
where a mass may diffuse as a whole or a part of it may chip off
\cite{ref_skmurthy}.  In this model, the distribution of mass at each
site undergoes a transition from an exponential to a power-law as the
overall density is increased. Beyond this critical density, the
power-law distribution is unchanged; the extra mass appears as an
infinite aggregate - a single large mass which is a finite fraction of
the total mass in the system. If the jumps to the left and right are
made asymmetric, this transition is not possible
\cite{ref_rajesh}. Although, in simulations, the transition as well as
the infinite aggregate is observed - this was attributed to
finite-size effects.

Further, if the rate of diffusion of the mass as a whole were not
constant, as in the above, but mass dependent $(\sim m^{-\gamma}$ with
$\gamma>0)$ \cite{ref_rajesh}, although the simulations show a
transition from the exponential mass distribution to a power-law with
an infinite aggregate as the density was increased, it turns out that
here too, this is only a finite-size effect and the model does not
show a true transition. 

Another example of an apparent phase transition reported in
\cite{ref_arndt}, the finite-size effects are discussed in
\cite{ref_kafri}.

\bibliographystyle{iopart-num}

\begin{thebibliography}{10}
\bibitem{ref_note1} For unequal rate of switching, $\alpha_{+}
  (\alpha_{-})$ being the rate of interconversion from a right (left)
  moving state to a left (right) moving state, we observe that the
  cluster only tends to drift along the lattice towards the left
  (right) for $\alpha_{+}<\alpha_{-}$ ($\alpha_{+}>\alpha_{-}$). In
  this work, we have taken them to be equal as our interest is in the
  cluster-size distributions and the jammed state.
\bibitem{ref_note2} As $Q$ is increased from $1$, the distribution of size
  of the clusters seems to change from an exponential to a
  stretched-exponential.
\end{thebibliography}
\providecommand{\newblock}{}

\end{document}